\input{aipcheck}

\documentclass[,final,numberedheadings]{aipproc}
\layoutstyle{6x9}

\newcommand{\Fermi}{\emph{Fermi}}
\newcommand{\FL}{\emph{Fermi} LAT}
\newcommand{\D}{$^{\circ}$}

\begin{document}
\title{Observations and Modeling of Gamma-ray Millisecond Pulsars seen with the \Fermi{} LAT}

\classification{95}
\keywords{gamma rays: observations, pulsars: millisecond}

\author{T. J. Johnson}{address={University of Maryland, Departments of Physics and Astronomy, College Park, MD 20742, USA, Tyrel.J.Johnson@nasa.gov}}
\author{C. Venter}{address={Centre for Space Research, North-West University, Potchefstroom 2520, South Africa}}
\author{A. K. Harding}{address={NASA Goddard Space Flight Center, Greenbelt, MD 20771, USA}}
\author{L. Guillemot}{address={Max-Planck-Institut f\"{u}r Radioastronomie, Auf den H\"{u}gel 69, 53121 Bonn, Germany.}}
\author{the \FL{} Collaboration and Pulsar Timing Consortium}{address={Across the world.}}

\begin{abstract}
We present a summary of gamma-ray millisecond pulsar (MSP) observations with the \Fermi{} Large Area Telescope. The radio and gamma-ray light curves of these MSPs have been modeled in the framework of the retarded vacuum dipole magnetic field. Likelihood fitting of the radio and gamma-ray light curves with geometric emission models allows us to give model-dependent confidence contours for the viewing geometry in these systems which are complementary to those from polarization measurements.
\end{abstract}

\maketitle

\section{Introduction}\label{intro}
Soon after the first millisecond pulsar (MSP) was discovered \cite{Backer82} there was much speculation regarding high-energy (HE, $\geq$0.1 GeV) MSP emission (e.g., \cite{Usov83}).  Most models assume that pulsar HE emission is the result of curvature radiation from electrons and/or positrons.  In these models the particles are accelerated either near the stellar surface (above the magnetic polar cap, e.g. \cite{PC}) or in the outer magnetosphere (up to the light cylinder) in narrow accelerating gaps with either a two-pole caustic (TPC, e.g. \cite{TPC}) or outer-gap (OG, e.g. \cite{OG}) geometry. The derived magnetic fields of MSPs, assuming dipole spin down, lie below the curvature radiation pair death line \cite{CRdl} on the P-$\dot{\rm P}$ diagram and thus narrow accelerating gaps were not expected.  This led to the development of the pair-starved polar cap (PSPC) model in which the full open field line volume above the polar cap is available to accelerate particles \cite{PSPC}.

Upper limits (3$\sigma$) were calculated for 19 MSPs in \emph{EGRET} data \cite{Fierro95} and a 4.9$\sigma$ pulsed detection using \emph{EGRET} data was reported for PSR J0218+4232 \cite{Kuiper00}.  Recently, 4.2$\sigma$ pulsations were detected from PSR B1821$-$24 in the globular cluster M28 using \emph{AGILE} data \cite{AGILEmsp}.  It was not clear if these singular cases were unique or if HE emission was common in MSPs until the launch of the \emph{Fermi Gamma-ray Space Telescope} (\Fermi{}).

\section{Millisecond Pulsars and \Fermi{}}\label{FermiMSPS}
The main instrument aboard \Fermi{} is the Large Area Telescope (LAT) \cite{LAT}, a pair-conversion telescope sensitive to gamma rays with energies from 0.02 to $>$300 GeV.  Using $\sim$8 months of data, significant HE pulsations from 8 MSPs were detected with the LAT \cite{MSPpop}.  Some MSP gamma-ray light curves showed sharp peaks reminiscent of what is seen in younger gamma-ray pulsars \cite{PSRcat}.  This implies that the HE emission processes in MSPs must be the same as those in younger gamma-ray pulsars.  In particular, MSPs are able to form narrow accelerating gaps despite being below the theoretical curvature radiation pair death line \cite{CRdl}.  This was supported by modeling of the gamma-ray and radio light curves for these 8 MSPs \cite{VHG} using TPC, OG, and PSPC models for the gamma-rays and a hollow-cone beam for the radio \cite{Story07}.  It was found that 6 were well fit with outer-magnetospheric models while only 2 required the PSPC model.

Pulsed gamma rays have also been detected from PSR J0034$-$0534 in which the gamma-ray and radio profiles are aligned \cite{J0034}, a phenomenon previously seen only in the Crab pulsar.  To match the observations, the radio emisson was modeled as significantly extended in altitude, near the light cylinder, and contained within the gamma-ray emisson region.

In addition to the pulsed detection of individual MSPs, gamma-ray point sources have been found consistent with 8 globular clusters \cite{GCpop} known or suspected of harboring many MSPs, which also display the characteristic pulsar spectrum, implying that the observed emission is from a combination of many MSPs.  Radio searches of unassociated \FL{} sources have revealed 23 new MSPs to date, some of which have now been seen to pulse in gamma-rays as well (e.g., \cite{Lucas}).

\section{Simulations and Likelihood Fitting}\label{SimsFit}
We have simulated radio and gamma-ray light curves following the procedures of \cite{VHG}, using the retarded vacuum dipole magnetic field geometry, with resolutions of 1\D{} in inclination angle ($\alpha$), the same in viewing angle ($\zeta$), and 0.05 in gap width (normalized to the polar cap opening angle).  Electrons are followed along the magnetic field lines out to a radial distance of 1.2 R$_{\rm LC}$ but not beyond a cylindrical distance of 0.95 R$_{\rm LC}$, where R$_{\rm LC}$ = cP/(2$\pi$).  We have also included the Lorentz transformation of the magnetic field from the inertial observer's frame to the co-rotating frame which has been advocated as necessary for self-consistency \cite{BS}.  To fit the light curves we have developed a Markov chain Monte Carlo (MCMC) maximum likelihood procedure which uses small-world chain steps \cite{SWC} and simulated annealing \cite{temp}.  An MCMC involves taking random steps in parameter space and accepting a step based on the likelihood ratio with respect to the previous step \cite{MH}.  It is necessary to balance the precision of the observed radio profiles against the simplistic cone-beam geometry we use and the best choice of radio uncertainty is under investigation. For the example presented in Section \ref{Results} we use an uncertainty equal to the average relative gamma-ray error, in the on-peak region, times the maximum radio value.

\section{Preliminary Results}\label{Results}
\begin{figure}[ht]
\includegraphics[width=0.5\textwidth]{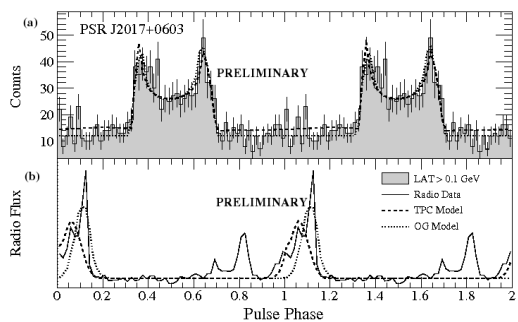}
\caption{ Data and best-fit light curves for PSR J2017+0603 with gamma-ray (a) and radio (b).  TPC models are the large dashed lines while OG models are the small dashed lines.}\label{lcs}
\end{figure}

Our MCMC algorithm has been applied to PSR J2017+0603, a 2.896 ms pulsar which is one of three MSPs discovered to date with the Nan\c{c}ay radio telescope in searches of \FL{} unassociated sources \cite{Cognard,Lucas}.  We find $\alpha$ = 16\D{} and $\zeta$ = 68\D{} with an infinitely thin gap for the TPC model and $\alpha$ = 36\D{} and $\zeta$ = 74\D{} with a gap width of 0.05 for the OG model, see Fig. \ref{lcs}.  For the TPC model, an infinitely thin gap is unphysical which implies that the best-fit value is less than our width resolution of 0.05.  The observed gamma-ray light curve is reproduced well by both models with the OG model slightly preferred by the likelihood.  Both models reproduce the correct radio-to-gamma phase lag but neither is successful at generating all of the observed radio features.  The complexity of the radio profile suggests that the emission may be from a region extended in altitude or that the radio beam shape is much more complex than a single-altitude cone.

\begin{figure}[ht]
\includegraphics[width=0.4\textwidth]{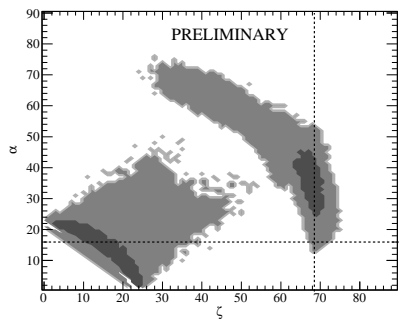}
\caption{Marginalized confidence contours for the TPC fit of PSR J2017+0603: within the darkest grey area is 39\% confidence, next is 68\%, and lightest grey area is 95\%.  The best-fit geometry is indicated by the dashed lines.}\label{conts}
\end{figure}

By examining the distribution of $\alpha$-$\zeta$ pairs in the output chains it is possible to obtain confidence contours in viewing geometry marginalized over the other parameters (Fig. \ref{conts} for the TPC model).  The resulting contours are not constraining but will improve with more gamma-ray data.  Examining the marginal distributions of the other fit parameters and $\alpha$ reveals very asymmetric probabilities which leads to the offset of the $\alpha$-$\zeta$ peak and the best-fit geometry.  There are currently no geometric constraints from radio polarization measurements of PSR J2017+0603 for comparison.

\section{Conclusions}\label{Conclusions}
MSPs have been established as a class of HE emitters and observations with the LAT have led to a reassessment of MSP HE emission models.   We have used geometric models to simulate the radio and gamma-ray light curves of PSR J2017+0603 and produced confidence contours in viewing geometry.  We plan to apply this technique to all LAT detected MSPs.

\begin{theacknowledgments}
The \Fermi{} LAT Collaboration acknowledges support from a number of agencies and institutes for both development and the operation of the LAT as well as scientific data analysis. These include NASA and DOE in the United States, CEA/Irfu and IN2P3/CNRS in France, ASI and INFN in Italy, MEXT, KEK, and JAXA in Japan, and the K. A. Wallenberg Foundation, the Swedish Research Council and the National Space Board in Sweden. Additional support from INAF in Italy and CNES in France for science analysis during the operations phase is also gratefully acknowledged.  This research was supported by the \Fermi{} GI program.
\end{theacknowledgments}

\end{document}